\documentclass[aps,superscriptaddress,twocolumn,floatfix,preprintnumbers,nofootinbib]{revtex4}
\usepackage{amsmath,multirow,graphicx,tabularx,epsfig}

\newcommand\one{\leavevmode\hbox{\small1\normalsize\kern-.33em1}}

\newcommand{\gev}{{\ensuremath\rm GeV}}

\def\slashchar#1{\setbox0=\hbox{$#1$}           
   \dimen0=\wd0                                 
   \setbox1=\hbox{/} \dimen1=\wd1               
   \ifdim\dimen0>\dimen1                        
      \rlap{\hbox to \dimen0{\hfil/\hfil}}      
      #1                                        
   \else                                        
      \rlap{\hbox to \dimen1{\hfil$#1$\hfil}}   
      /                                         
   \fi}

\def\eg{{\sl e.g.} \,}

\def\etal{{\sl et al} \,}

\newcommand{\be}{\begin{eqnarray*}}
\newcommand{\ee}{\end{eqnarray*}}

\newcommand{\bee}{\begin{eqnarray}}
\newcommand{\eee}{\end{eqnarray}}
\newcommand{\beeq}{\begin{equation}}
\newcommand{\eeeq}{\end{equation}}

\begin{document}



\title{Higgs Couplings after the Discovery}

\author{Tilman Plehn}
\affiliation{Institut f\"ur Theoretische Physik, Universit\"at Heidelberg, Germany}

\author{Michael Rauch}
\affiliation{Institut f\"ur Theoretische Physik, Karlsruhe Institute of Technology (KIT), Germany}

\begin{abstract}
 Following the ATLAS and CMS analyses presented around ICHEP 2012 we
 determine the individual Higgs couplings. The new data allow us to
 specifically test the effective coupling to photons. We find no
 significant deviation from the Standard Model in any of the Higgs
 couplings.
\end{abstract}

\maketitle


A resonance peak consistent with a scalar Higgs boson~\cite{higgs} has
been discovered by ATLAS~\cite{atlas} and CMS~\cite{cms}.  It
completes the Standard Model of elementary particles and establishes the
fundamental concept of gauge theories.

In its Standard Model form the Higgs boson couples to all particles
proportional to their masses~\cite{abdel,spirix,lecture}. Beyond the
Standard Model, Higgs analyses probe new physics orthogonally to
direct searches~\cite{bsm_review}. Induced dimension-five couplings to
gluons and to photons are among the key parameters for Higgs analyses
at the LHC.  In addition, new physics can couple to the Higgs sector
via renormalizable dimension-four operators~\cite{portal}.  Similarly,
strongly interacting models will alter all Higgs couplings in a way
which reflects the underlying
theory~\cite{us,sfitter_higgs,duehrssen,others,rome,new_ops,lc}.\medskip

\underline{{\sc SFitter} Higgs analyses} --- the general Higgs
coupling analysis and its first application on data is comprehensively
documented in
Refs.~\cite{us,sfitter_higgs,sfitter}.
\footnote{Continuous Updates of some figures in this letter can be
  found under \url{www.thphys.uni-heidelberg.de/~plehn}} All
tree-level Higgs couplings and their ratios are parameterized as
\begin{alignat}{9}
g_{xxH} &\equiv g_x  = 
\left( 1 + \Delta_x \right) \;
g_x^\text{SM} 
\notag \\
\frac{g_{xxH}}{g_{yyH}} &\equiv
\frac{g_x}{g_y} = 
\left( 1 + \Delta_{x/y} \right) \; 
\left( \frac{g_x}{g_y} \right)^\text{SM} \; .
\label{eq:delta}
\end{alignat}
For the loop-induced Higgs-photon coupling this means
\begin{alignat}{9}
g_{\gamma\gamma H} &\equiv g_{\gamma}  = 
\left( 1 + \Delta_\gamma^\text{SM} + \Delta_\gamma \right) \;
g_\gamma^\text{SM} \ .
\label{eq:deltagamma}
\end{alignat}
The Standard Model value is given by the bottom, top, and $W$ loops.
$\Delta_\gamma^\text{SM}$ contains the measured direct couplings to
all Standard Model particles, $\Delta_\gamma$ parameterizes additional
contributions to the effective vertex.  

All underlying operators we assume to be Standard-Model like.  A
dedicated analysis of the `Higgs' quantum numbers requires a detailed
study of angular correlations in $H \to ZZ$ decays~\cite{lookalikes}
or weak boson fusion Higgs production~\cite{wbf_coup}.\medskip

The Higgs width cannot be measured independently at the LHC, but it
enters every observable event rate as an over-all normalization. We
consistently assume~\cite{hdecay}
\begin{alignat}{9}
\Gamma_\text{tot} = \sum_\text{obs} \; \Gamma_x(g_x) 
+ \text{2nd generation} < 2\,\gev
\; .
\label{eq:width}
\end{alignat}
At the LHC we cannot observe $g_c$. To avoid systematic offsets in our
results we link all second-generation Yukawa couplings to their
third-generation counter parts, \eg $g_c = m_c/m_t \times
g_t^\text{SM} (1 + \Delta_t)$.\medskip

Experimental and theoretical
uncertainties~\cite{gf_rate,wbf_rate,xs_group} enter our fit including
full correlations. Theoretical uncertainties we include in the
centrally flat {\sc Rfit} scheme~\cite{rfit}.  All background rates,
efficiencies and experimental uncertainties are extracted from ATLAS
and CMS publications~\cite{atlas,cms}.  Based on all this we compute
an exclusive log-likelihood map of the Higgs parameter
space. Individual distributions are defined through a profile
likelihood. The best-fitting points we identify using Cooling Markov
Chains~\cite{sfitter_higgs} combined with {\sc Minuit}, their 68\%
confidence intervals require $5000$ toy measurements.\medskip

\begin{figure*}[t]
\includegraphics[width=0.38\textwidth]{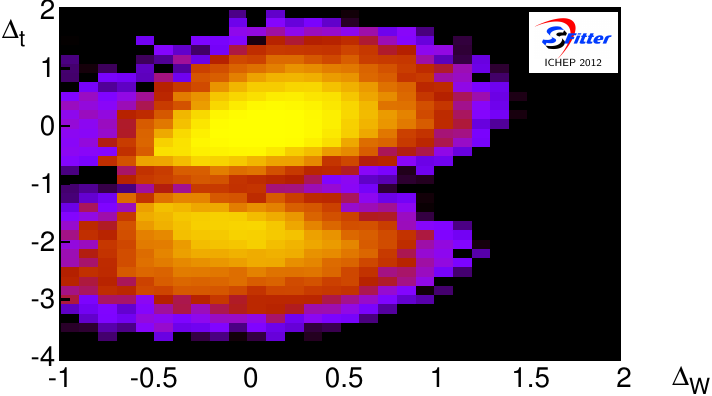}
\hspace*{0.05\textwidth}
\includegraphics[width=0.38\textwidth]{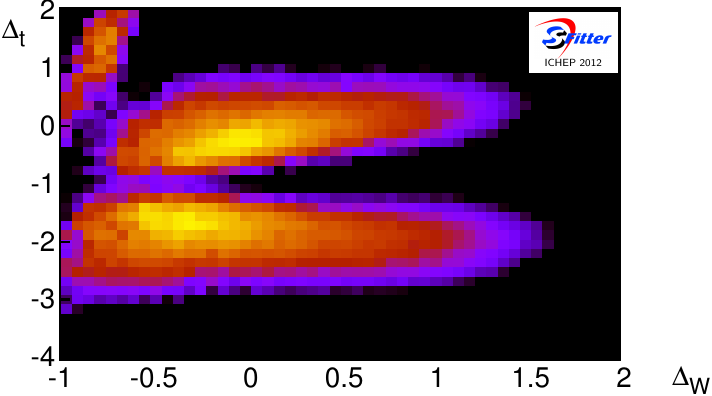}
\hspace*{0\textwidth}
\raisebox{-0.5pt}{\includegraphics[width=0.049\textwidth]{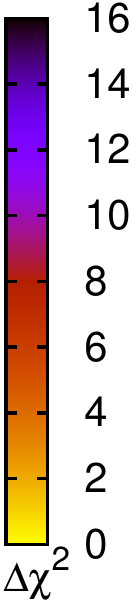}}
\vspace*{-3mm}
\caption{$\Delta_W$ vs $\Delta_t$ for the expected SM measurements
  (left) and the actual measurements (right), assuming $m_H =
  126$~GeV.}
\label{fig:markov}
\end{figure*}

\underline{Global picture} --- the analysis of the 7~TeV and 8~TeV
Higgs channels closely follows Ref.~\cite{us}. In this update we
include all recently published ATLAS~\cite{atlas} and CMS~\cite{cms}
analyses, including the recent $H \to WW$ ATLAS results.

Before we show the measured individual Higgs couplings we briefly
discuss the global structure of the log-likelihood map, focusing on
correlations. Because we now allow for an independent $g_\gamma$
variation we are most interested in the correlation between the Higgs
couplings to top quarks and to $W$-bosons.

In the left panel of Fig.~\ref{fig:markov} we see the expected
correlation of these two couplings in the absence of an additional
parameter $\Delta_\gamma$. This log-likelihood map is based on the
proper errors of the 2011 and 2012 analyses, but assuming Standard
Model central values. Two solutions arise from a sign change in the
top Yukawa coupling, $\Delta_t \sim -2$. Only $g_W$ we keep positive
without loss of generality. The preference for positive $g_t$
indicates that the degeneracy of the two solutions is broken by the
amplitude-level interference between the top and $W$ loops in the
effective $g_\gamma$ computation.

In the right panel of Fig.~\ref{fig:markov} we show the same
distribution based on data. The two expected solutions indeed
appear, but a third scenario features around $\Delta_W \sim -1$ and
$\Delta_t \sim 1.5$. In this case the $W$-boson decouples from the
Higgs, as it is still allowed within errors. The usually smaller top
loop now induces the effective vertex on its own, requiring a slightly
enhanced Yukawa coupling.\medskip

In contrast to the earlier analysis of Ref.~\cite{us} we do not
attempt to constrain the parameter space to Standard-Model like
solutions. This means that secondary solutions will appear with a
competitive quality of fit. For example, including all Standard Model
Higgs couplings, but no free coupling to photons we find three equally
likely points:

\setlength{\tabcolsep}{5pt}
\begin{small} \begin{center} \begin{tabular}{rrrrr|r@{/}l}
\hline
$\Delta_W$ & $\Delta_Z$ & $\Delta_t$ & $\Delta_b$ & $\Delta_\tau$ & $\chi^2$ & d.o.f. \\\hline
-0.03 & -0.02 & -0.25 & -0.25 & -0.90 & 27.7 & 49 \\
-0.05 & -0.04 & -0.34 & {\it -1.73} & -0.70 & 27.6 & 49 \\
-0.29 & -0.09 & {\it -1.65} & -0.32 & -0.70 & 27.7 & 49 \\
\hline
\end{tabular} \end{center} \end{small}

In the second and third line the bottom and top Yukawa
coupling, respectively, have changed sign. As expected, we cannot
distinguish such alternative scenarios with the current data.\medskip

\underline{Local picture} --- from the exclusive log-likelihood maps we
can extract individual Higgs couplings. We are directly sensitive to
$\Delta_{W,Z,\tau}$ and can extract $\Delta_t$ from the effective
photon and gluon couplings as well as $\Delta_b$ from the total
width. As we will see below, we can even constrain an additional free
parameter $\Delta_\gamma$.\medskip

\begin{figure}[b]
\includegraphics[width=0.42\textwidth]{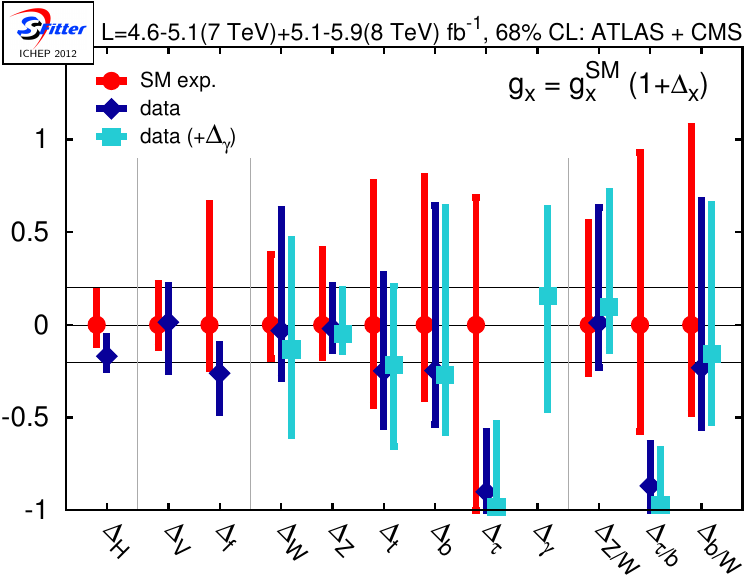}
\caption{Results based on 2011 and 2012 data, for the SM signal
  expectation and for the data ($m_H = 126$~GeV). We also show the
  form factor result $\Delta_H$ and universal fermion and boson
  couplings $\Delta_{V,f}$.  The band indicates a $\pm 20\%$
  variation.}
\label{fig:sm8}
\end{figure}

Of course, extracting any smaller number of model parameters is
technically easier and will lead to smaller error bars. For example,
we can test a hypothetical universal form factor of all tree-level
Higgs couplings,
\begin{alignat}{5}
  \Delta_x \equiv \Delta_H  \qquad \text{for all} \; x \; .
\end{alignat}
In Fig.~\ref{fig:sm8} we show the expected and observed central value
and error bar on this form factor. Such a form factor is barely
consistent with the Standard Model value $\Delta_H = 0$. Its low
central value is a result of all three third-generation Yukawa
couplings tending towards smaller values. Quoting this result we need
to keep in mind that it is only sensible if all individual $\Delta_x$
are consistent.\medskip

Two-parameter fits, on the gauge coupling side, are motivated by
electroweak precision data. In the absence of new physics signals
these measurements point towards $\Delta_W = \Delta_Z$. We define
\begin{alignat}{5}
  \Delta_W = \Delta_Z &\equiv \Delta_V \notag \\
  \Delta_b = \Delta_t = \Delta_\tau &\equiv \Delta_f \; .
\end{alignat}
While the vector boson coupling in Fig.~\ref{fig:sm8} is measured in
complete agreement with the Standard Model the Yukawa couplings have
consistently low best-fit values. However, within the uncertainties
this is no problem.\medskip

Obviously, what we really want to extract (if possible) is the set of
all couplings individually. Figure~\ref{fig:sm8} shows the central
coupling values for $W$ and $Z$-bosons as well as the third-generation
fermions.

Comparing the expected to the observed uncertainties we see that the
two massive gauge couplings are extracted very well after including
the 8~TeV data. The indirectly measured top and bottom Yukawa
couplings come out slightly low, but agree with the Standard Model
expectations within relatively large error bars.  Those are due to
their indirect determination.  A tau Yukawa coupling is not
experimentally established yet. For example comparing the measured
value of $\Delta_b$ with the ratio $\Delta_{b/W}$ we see no
significant improvement. The ATLAS and CMS measurements are still
largely statistics limited, so forming ratios does not help.

As widely discussed in the literature, the observed number
of Higgs events decaying to photons is slightly larger than the
Standard Model expectation. In Fig.~\ref{fig:sm8} we see that
without a free Higgs coupling to photons the best fit resides around
$\Delta_W \sim 0$. The central point for the top Yukawa is $\Delta_t
\sim -0.25$, enhancing the Higgs branching ratio to photons, but
reducing the production cross section from gluon fusion. However,
the central value of $\Delta_b$ is half a standard deviation below
the Standard Model expectation, corresponding to a slightly
increased number of Higgs events in the two-photon channel.\medskip

\underline{Free Higgs-photon coupling} --- finally, in
Fig.~\ref{fig:sm8} we show the fit to all Standard Model couplings
plus a free shift in the higher-dimensional photon-Higgs coupling. The
$\Delta_W$ measurement now mainly relies on the observed $H \to WW$
decays, without the additional information from the photon decay mode;
its central value moves down by 13\%, well within its
uncertainty. The top Yukawa is still extracted from the
Higgs coupling to gluons and comes out slightly low. Similarly, the
bottom Yukawa contributing to the total width is 25\% below the
Standard Model expectation. These two effects from the Yukawa
coupling roughly cancel each other, but the reduced value of
$\Delta_W$ is compensated by the effective Higgs-photon
coupling. Note that all shifts are within one standard deviation for
the respective couplings. For this additional higher-dimensional
coupling as defined in Eq.\eqref{eq:deltagamma} we find
\begin{alignat}{5}
\Delta_\gamma = 
0.16 + 0.47 - 0.61 \quad (68\%~\text{CL}) \; .
\end{alignat}
For a switched sign of $g_t$, which
yields the same log-likelihood value, this central value moves to
$\Delta_\gamma = - 0.22$. Hence, we find no evidence for an enhanced
Higgs coupling to photons; this is in apparent conflict with the ATLAS
and CMS results. The reason is that we consistently use the best-fit
values for all other couplings ($1 + \Delta_\gamma^\text{SM}$) in the
comparison between theory and experiment for the $H \to \gamma \gamma$
channels, while ATLAS and CMS use the Standard Model input.\medskip

Starting from Fig.~\ref{fig:sm8} we see the next steps: improved
statistics in the weak-boson-fusion channel will improve the $g_\tau$
measurement~\cite{wbf_tau}.  To be able to test the induced
Higgs-gluon coupling we need a direct determination of the top Yukawa
coupling in $t\bar{t}H$ production.  Similarly, to probe contributions
to the Higgs width a direct measurement of the bottom Yukawa coupling
is mandatory.  Both will hugely benefit from Higgs and top tagging
analyses~\cite{boosted}.\medskip

\underline{Outlook} --- in this updated Higgs coupling analysis
following Ref.~\cite{us} we show that all ATLAS and CMS Higgs
measurements are unfortunately well explained by a Standard Model
Higgs boson. In addition to some toy models we determine all Standard
Model Higgs couplings individually, including an independent Higgs
coupling to photons. None of the measured couplings deviates from its
Standard Model values significantly.\medskip

\underline{Acknowledgments} --- we would like to thank Dirk Zerwas,
Markus Klute, and Remi Lafaye for an absolutely great collaboration on
{\sc SFitter-Higgs}.



\begin{thebibliography}{99}

\bibitem{higgs}
 P.~W.~Higgs,
  Phys.\ Lett.\  {\bf 12}, 132 (1964);
 P.~W.~Higgs,
  Phys.\ Rev.\ Lett.\  {\bf 13}, 508 (1964);
 F.~Englert and R.~Brout,
  Phys.\ Rev.\ Lett.\  {\bf 13}, 321 (1964).

\bibitem{atlas}
The ATLAS Colloboration,
  arXiv:1207.0319;
Phys.\ Rev.\ Lett.\  {\bf 108}, 111803 (2012);
Phys.\ Lett.\ B {\bf 710}, 383 (2012);
arXiv:1206.0756;
arXiv:1206.5971;
arXiv:1207.0210;
ATLAS-CONF-2012-093; 
ATLAS-CONF-2012-091;
ATLAS-CONF-2012-092;
ATLAS-CONF-2012-098.

\bibitem{cms}
The CMS Collaboration,
S.~Chatrchyan {\it et al.}  [CMS Collaboration],
  Phys.\ Lett.\ B {\bf 710} (2012) 26;
Phys.\ Lett.\ B {\bf 710}, 91 (2012);
Phys.\  Rev.\  Lett.\  108, {\bf 111804} (2012);
Phys.\ Lett.\ B {\bf 710}, 403 (2012);
Phys.\ Lett.\ B {\bf 713}, 68 (2012);
Phys.\ Lett.\ B {\bf 710}, 284 (2012);
CMS-PAS-HIG-12-025;
CMS-PAS-HIG-12-020;
CMS-PAS-HIG-12-015;
CMS-PAS-HIG-12-016;
CMS-PAS-HIG-12-017;
CMS-PAS-HIG-12-018;
CMS-PAS-HIG-12-019.

\bibitem{abdel}
 A.~Djouadi,
  Phys.\ Rept.\  {\bf 457}, 1 (2008).
 
\bibitem{spirix}
 M.~Spira,
  Fortsch.\ Phys.\  {\bf 46}, 203 (1998).

\bibitem{lecture}
 for a pedagogical introduction see 
  T.~Plehn,
  Lect.\ Notes Phys.\  {\bf 844}, 1 (2012)
  [arXiv:0910.4182].

\bibitem{bsm_review} 
 D.~E.~Morrissey, T.~Plehn and T.~M.~P.~Tait,
  arXiv:0912.3259;
 P.~Nath, B.~D.~Nelson, H.~Davoudiasl, B.~Dutta, D.~Feldman, Z.~Liu, T.~Han and P.~Langacker {\it et al.},
  Nucl.\ Phys.\ Proc.\ Suppl.\  {\bf 200-202}, 185 (2010).

\bibitem{portal}
 C.~Englert, T.~Plehn, M.~Rauch, D.~Zerwas and P.~M.~Zerwas,
  Phys.\ Lett.\ B {\bf 707}, 512 (2012),
  and references therein.

\bibitem{us} 
 M.~Klute, R.~Lafaye, T.~Plehn, M.~Rauch and D.~Zerwas,
  Phys.\ Rev.\ Lett.\ in print, arXiv:1205.2699 [hep-ph].

\bibitem{sfitter_higgs}
 R.~Lafaye, T.~Plehn, M.~Rauch, D.~Zerwas and M.~D\"uhrssen,
  JHEP {\bf 0908}, 009 (2009),
  and references therein.

\bibitem{duehrssen}
 M.~D\"uhrssen, 
 ATL-PHYS-2002-030;
 M.~D\"uhrssen \etal 
  Phys.\ Rev.\ D {\bf 70}, 113009 (2004);
 for an early analysis see also 
 D.~Zeppenfeld, R.~Kinnunen, A.~Nikitenko and E.~Richter-Was,
  Phys.\ Rev.\ D {\bf 62}, 013009 (2000).

\bibitem{others} 
 for more or less constrained Higgs analyses see \eg
  F.~Bonnet, M.~B.~Gavela, T.~Ota and W.~Winter,
  Phys.\ Rev.\ D {\bf 85}, 035016 (2012);
 D.~Carmi, A.~Falkowski, E.~Kuflik and T.~Volansky,
  arXiv:1202.3144
  and arXiv:1206.4201;
 D.~Carmi, A.~Falkowski, E.~Kuflik, T.~Volansky and J.~Zupan,
  arXiv:1207.1718;
 P.~P.~Giardino, K.~Kannike, M.~Raidal and A.~Strumia,
  JHEP {\bf 1206}, 117 (2012)
  and arXiv:1207.1347;
 J.~Ellis and T.~You,
  arXiv:1204.0464
  and arXiv:1207.1693;
 J.~R.~Espinosa, C.~Grojean, M.~M\"uhlleitner and M.~Trott,
  JHEP {\bf 1205}, 097 (2012);
 J.~R.~Espinosa, M.~Muhlleitner, C.~Grojean and M.~Trott,
  arXiv:1205.6790;
 J.~R.~Espinosa, C.~Grojean, M.~Muhlleitner and M.~Trott,
  arXiv:1207.1717.

\bibitem{rome}
 A.~Azatov, R.~Contino and J.~Galloway,
  arXiv:1202.3415;
 A.~Azatov, R.~Contino, D.~Del Re, J.~Galloway, M.~Grassi and S.~Rahatlou,
  arXiv:1204.4817.


\bibitem{new_ops}
 S.~Dawson and E.~Furlan,
  arXiv:1205.4733;
 I.~Low, J.~Lykken and G.~Shaughnessy,
  arXiv:1207.1093;
 T.~Corbett, O.~J.~P.~Eboli, J.~Gonzalez-Fraile and M.~C.~Gonzalez-Garcia,
  arXiv:1207.1344;
 S.~Banerjee, S.~Mukhopadhyay and B.~Mukhopadhyaya,
  arXiv:1207.3588;
 F.~Bonnet, T.~Ota, M.~Rauch and W.~Winter,
  arXiv:1207.4599.

\bibitem{lc}
 for a recent linear collider perspective see \eg
 M.~E.~Peskin,
  arXiv:1207.2516.

\bibitem{sfitter} 
 R.~Lafaye, T.~Plehn, M.~Rauch and D.~Zerwas,
  Eur.\ Phys.\ J.\ C {\bf 54}, 617 (2008).

\bibitem{lookalikes}
 Y.~Gao, A.~V.~Gritsan, Z.~Guo, K.~Melnikov, M.~Schulze and N.~V.~Tran,
  Phys.\ Rev.\ D {\bf 81}, 075022 (2010);
 A.~De Rujula, J.~Lykken, M.~Pierini, C.~Rogan and M.~Spiropulu,
  Phys.\ Rev.\ D {\bf 82}, 013003 (2010);
  J.~Ellis and D.~S.~Hwang,
  arXiv:1202.6660 [hep-ph];
  D.~Zeppenfeld {\it et al.},
  arXiv:1207.4975 [hep-ph].

\bibitem{wbf_coup}
 see \eg
 T.~Plehn, D.~L.~Rainwater and D.~Zeppenfeld,
  Phys.\ Rev.\ Lett.\  {\bf 88}, 051801 (2002);
 K.~Hagiwara, Q.~Li and K.~Mawatari,
  JHEP {\bf 0907}, 101 (2009);
 C.~Englert, M.~Spannowsky and M.~Takeuchi,
  arXiv:1203.5788.

\bibitem{hdecay}
  A.~Djouadi, J.~Kalinowski and M.~Spira,
  Comput.\ Phys.\ Commun.\  {\bf 108}, 56 (1998).

\bibitem{gf_rate}
 H.~M.~Georgi, S.~L.~Glashow, M.~E.~Machacek and D.~V.~Nanopoulos,
  Phys.\ Rev.\ Lett.\  {\bf 40}, 692 (1978);
 S.~Dawson,
  Nucl.\ Phys.\ B {\bf 359}, 283 (1991);
 M.~Spira, A.~Djouadi, D.~Graudenz and P.~M.~Zerwas,
  Nucl.\ Phys.\ B {\bf 453}, 17 (1995);
 R.~V.~Harlander and W.~B.~Kilgore,
  Phys.\ Rev.\ Lett.\  {\bf 88}, 201801 (2002);
 C.~Anastasiou and K.~Melnikov,
  Nucl.\ Phys.\ B {\bf 646}, 220 (2002);
 V.~Ravindran, J.~Smith and W.~L.~van Neerven,
  Nucl.\ Phys.\ B {\bf 665}, 325 (2003);
 V.~Ahrens, T.~Becher, M.~Neubert and L.~L.~Yang,
  Eur.\ Phys.\ J.\ C {\bf 62}, 333 (2009).

\bibitem{wbf_rate} 
  M.~Ciccolini, A.~Denner and S.~Dittmaier,
  Phys.\ Rev.\ D {\bf 77}, 013002 (2008);
 K.~Arnold {\it et al.},
  arXiv:1107.4038.

\bibitem{xs_group}
 S.~Dittmaier {\it et al.}  
  arXiv:1101.0593.

\bibitem{rfit}
 A.~H\"ocker, H.~Lacker, S.~Laplace and F.~Le Diberder,
  Eur.\ Phys.\ J.\  C {\bf 21}, 225 (2001).

\bibitem{wbf_tau}
 D.~L.~Rainwater, D.~Zeppenfeld and K.~Hagiwara,
  Phys.\ Rev.\  D {\bf 59}, 014037 (1999);
 T.~Plehn, D.~L.~Rainwater and D.~Zeppenfeld,
  Phys.\ Rev.\  D {\bf 61}, 093005 (2000).

\bibitem{boosted}
  J.~M.~Butterworth, A.~R.~Davison, M.~Rubin, G.~P.~Salam,
  Phys.\ Rev.\ Lett.\  {\bf 100}, 242001 (2008);
 T.~Plehn, G.~P.~Salam and M.~Spannowsky,
  Phys.\ Rev.\ Lett.\  {\bf 104}, 111801 (2010).


\end{thebibliography}
\end{document}